\documentstyle[11pt]{article}
\newcommand{\be}{\begin{equation}}
\newcommand{\ee}{\end{equation}}
\newcommand{\bea}{\begin{eqnarray}}
\newcommand{\eea}{\end{eqnarray}}

\newcommand{\m}{\mu}
\newcommand{\n}{\nu}
\begin{document}
\begin{center}
\hspace*{10cm} TUM-HEP-256/96\\
\hspace*{10cm} LMU-TPW 96-25\\
\hspace{10cm} hep-th/9610060\\
\vspace*{.5cm}

{\LARGE ZERO--BRANES IN  2+1 DIMENSIONS\footnote{
Talk given by A. Kehagias
in ``Gauge Theories, Applied Supersymmetry and Quantum Gravity",
Imperial College, July 1996}$^,$\footnote{Work supported
by the EC programm SC1-CT91-0729 and the European Commission 
TMR programmes ERBFMRX-CT96-0045
and ERBFMRX-CT96-0090.}}\\

\vspace{.3cm}

{\large Stefan F\"orste\footnote{ 
Supported by GIF--German Israeli Foundation for Scientific Reseach. E-Mail:Stefan.Foerste@physik.uni-muenchen.de}

{\it { Sektion Physik,
Universit\"at M\"unchen\\
\vspace{-.2cm}
Theresienstra\ss e 37, 80333 M\"unchen\\
Germany}}\\
\vspace{.1cm}

and
\vspace{.1cm}

Alexandros Kehagias
\footnote{Supported by the Alexander von Humboldt Stiftung. 
E-Mail: kehagias@physik.tu-muenchen.de}

{\it { Physik Department \\
\vspace{-.2cm}
Technische Universit\"at M\"unchen\\
D-85748 Garching, Germany}}}\\
\end{center}
\vspace{.5cm}

\begin{center}
{\large Abstract}
\end{center}
\vspace{.2cm}

We discuss  zero-brane solutions  in three-dimensional
Minkowski space-time 
annihilated by half of the supersymmetries. The other half of the
supersymmetries which should generate the supersymmetric multiplet
are ill-defined and lead to a non bose-fermi degenerate solitonic
spectrum.

\section{Introduction}
There exists a  boson-fermion degeneracy of the supermultiplet in
supersymmetric theories.
However, there are cases where such degeneracies do not
appear. A U(1) gauge theory for example in 2+1 dimensions has
Nielsen-Olesen vortex solutions \cite{1} annihilated by half of the
supersymmetries~\cite{2,3}.
The other half generates zero modes which form a
supermultiplet of the unbroken N=1 supersymmetry.
This happens as
long as one restricts himself to global supersymmetry. In the
supergravity context, Nielsen-Olesen 
vortex solutions which break  half of
the supersymmetries also exist.
 However, the other half of the supersymmetries now 
generates non-normalizable zero modes
and thus the latter do not appear in the physical
Hilbert space~\cite{3}.
As a result, the U(1) vortex solutions do not have fermionic
partners which are neccesary to form  a supermultiplet of the
unbroken supersymmetry. Thus, one finds a solitonic spectrum
without bose-fermi degeneracies.
The reason of this peculiarity 
is the conical structure of space-time in 2+1
dimensions~\cite{4,5}.
 It is known that away from the sources the space-time is
flat and there is a deficit angle at spatial infinity. The deficit
angle does not allow the existence of Killing spinors~\cite{5',6}  and,
consequently, there are no states which carry fermionic supercharge.
The latter cannot be consistently defined.
In the presence of gauge fields, Killing spinors may exist as
in the U(1) case mentioned above and states with fermionic supercharge
do exist. However, they are not normalizable.

Here we will examine a  similar case, namely, we will show that the
solitonic spectrum of zero-branes in 2+1 dimensions  has no bose-fermi
degeneracies.
These considerations are also  related to Witten's
observation~\cite{w} that in
2+1 dimensions the vacuum can have exactly zero cosmological constant
because of supersymmetry while excited states may not come in
boson-fermion pairs~\cite{3,nishino}.

\section{Global supercharges}

Charges are defined as spatial integrals of the time component of a conserved
current. However, in many cases there are currents which 
satisfy  covariant conservation laws due to some local symmetry. 
One may recall for example  the energy--momentum tensor in
general relativity or the Yang-Mills current in gauge theories. In the first 
case  an ordinarily conserved energy--momentum tensor 
may be constructed if there exist Killing
vectors~\cite{deserab}.  
For an asymptotically
D+1-dimensional Minkowski space, there exist  $(D+1)(D+2)/2$ Killing vectors
which give rise to  $(D+1)(D+2)/2$ conserved charges~\cite{teit}
\bea
P_0&=&\oint_{r\rightarrow\infty}dS_i(g_{ij,j}-g_{jj,i}), \,\\
P^j&=&\oint_{r\rightarrow\infty}dS_i \pi^{ij}\, ,\\
J^{ik}&=&\oint_{r\rightarrow\infty}dS_k(x^i\pi^{jk}-x^k\pi^{ij})\, ,
\\
J_{0k}&=&\oint_{r\rightarrow\infty}dS_i\left(x^k(g_{ij,j}-g_{jj,i})
-g_{ik}+g_{jj}\delta_{ik}\right)\,  . \label{charges}
\eea
One may prove that they satisfy the Poincar\'e algebra  while enlarged 
with the
spinor charges they all form the supersymetric algebra.

In order to define the spinor charges~\cite{deserab}
 let us consider the vector-spinor
density  in $D+1$ dimensions 
\be
j^\m= \Gamma^{\m\n\lambda}D_\n\psi_\lambda \, , \label{RS}
\ee
where as usual, $\Gamma^{\m\n\lambda}=\Gamma^{[\mu}\Gamma^\n
\Gamma^{\lambda]}$, $(\m,\n=0,\cdots,D\, , m,n=1,\cdots,D)$. 
The current $j^\m$ is not conserved
but rather,  as the Yang-Mills current and the energy--momentum tensor,
 it is covariantly constant $D_\m j^\m=0$. However, one may constructs a
conserved quantity out of $j^\m$ if the space-time admits 
a covariantly constant (Killing) spinor $\epsilon$, i.e., 
\be
D_\m \epsilon=0\, .\label{cs}
\ee
Then, one may form the vector $J^\m=\bar{\epsilon}j^\m$ which may be written as
\be
J^\m=\partial_\n(\bar{\epsilon}
\Gamma^{\mu\nu\kappa}\psi_\kappa)\, , \label{J}
\ee
and thus it satisfies $\partial_\m J^\m=0$. As a result, the spinor charge
\be
Q(\epsilon)=\int d^Dx\sqrt{g^{(D)}}J^0=\oint_{\infty} dS_i\bar{\epsilon}
\Gamma^{0ij} \psi_j \, , \label{qq}
\ee
is a conserved quantity.  One may observe that in order to have a non-zero
spinor charge, the covariantly constant spinor $\epsilon$ must approach a
constant spinor $\epsilon_0$ at spatial infinity.
In this case, there exist at
most $2^{[D+1/2]}$ spinor charges, one for each non-zero 
component of $\epsilon_0$. 
However, global charges may not always be possible to be defined in
gravitational  as well as in gauge theories. One
may recall for example that global colour does not exist in a monopole
background~\cite{color} and
momentum cannot be defined in the gravitational field
of a massive object in 2+1 dimensions~\cite{4}.
In our case, the lack of global
charges may be traced to the absence of the associated Killing vectors
and spinors which are essential for the definition of such charges as we will 
see below.

\section{Global supercharges in 2+1 dimensions}

Let us now consider the rather special  2+1-dimensional case.
Here, the space is flat out of the sources since the
Riemann tensor is
completely determined  by the Einstein tensor. For localized
sources, there exists a
peculiar conical structure at spatial infinity due to a deficit angle.
This leads to
breakdown of the Poincar\'e symmetry
at spatial infinity althought the space-time
is flat there~\cite{deser}. Asymptotically one
should expect, in view of the local flatness, the Killing vector
\be
\xi^\mu=a^\mu+\omega^\m_\n x^\n \, , \label{kil}
\ee
to generate space-time rotations and translations. If a deficit
angle $\beta$ is present, however,  one has to
identify $x^i=\Omega^i_j(\beta)x^j$ where
$\Omega(\beta)$ is an SO(2) rotation matrix with parameter
$\beta$~\cite{5}.
Then, it  may easily  be  proved that this condition
breaks some space-time symmetries and the only surviving ones are
the time-translations and space rotations. The other
symmetries fail to satisfy the continuity conditions along the seams.
As a result, the only conserved charges at spatial infinity
are the energy and
the angular momentum while the ordinary momenta do not exist.
 One of course expects  energy and
momenta to form a 3-vector under Lorentz
transformations and by boosting energy states to obtain states with
non-zero momentum. However, this is not possible since  the
momentum of the boosted states diverges
which  simply reflects the fact that the
asymptotic Poincar\'e invariance has been broken~\cite{deser}.
On the other hand, the spinor
charges are given by
\be
Q(\epsilon)=\oint_{r\rightarrow\infty} dS_j\bar{\epsilon}
\varepsilon^{ij}\psi_j\, ,\label{q3}
\ee
and since there are no covariantly constant spinors $\epsilon$ in conical
space-times they do not  exist as well.
The reason of the non-existence of Killing spinors
 is the non half-integral phase
acquired by  $\epsilon$ as it goes around a circle at
infinity~\cite{5'}.

Covariant constant spinors and, consequently spinor charges,
may be defined if one couples the spinors
to a U(1) gauge field. Then Killing spinors may
exist if  the non half-integral phase  acquired by the spinors as we parallel
transport them around a circle at infinity is 
canceled by an Aharonov-Bohm phase 
due to the U(1) gauge field. This  kind of situation 
has been considered in Ref. 3.
It is similar to  the generalized
spin structures on manifolds with
no ordinary spin structure. 
It is also possible that the U(1) gauge field cancels exactly the spin
connection \cite{a,b}. 
In this case, covariantly constant spinors exist as well and these
are the only  cases we know where global supercharges may
consistently be defined on a conical space-time.

The bosonic part of the action
in 2+1 dimensions where only the dilaton $\phi$ and an axion
$\alpha$ are not vanishing is
\be
I=\int d^3x\sqrt{-g}\left(R-\frac{1}{2}
\frac{\partial_\m S\partial^\m \bar{S}}
{(S-\bar{S})^2} \right) \, . \label{d3}
\ee
It describes a SL(2,R)/U(1) $\sigma$-model coupled to gravity
and  can be obtained by wrapping the seven-brane~\cite{b} of type IIB
around a six-cycle. It can also be obtained by compactifying a five-dimensional
theory on a two torus of constant volume~\cite{g}. 
The complex scalar S is given in terms of the
dilaton and the axion as $S=\alpha+ie^{-\phi}$ and belongs to the
upper half plane ($ImS>0$).
The action (\ref{d3}) is invariant under the following N=2
supersymmetry transformations~\cite{d,b} on a pure bosonic background
\bea
\delta\lambda&=&-\frac{1}{S_2}\left(\frac{\bar{S}-i}{S+i}\right)
\gamma^\m\partial_\m S
\epsilon^{\ast}\, , \nonumber \\
\delta \psi_\m&=&D_\m \epsilon\, , \label{sup}
\eea
where $\epsilon=\epsilon_1+i\epsilon_2$ is a complex spinor and 
\be
D_\m=\partial_\m+\frac{1}{4}
\omega_{\m ab}\gamma^a\gamma^b-\frac{i}{2}Q_\m \, ,
\ee
is a covariant derivative containing the  
spin connection as well as    the composite U(1) gauge field
\be
Q_\m=-\frac{1}{S_2}\left[\left(\frac{S-i}{\bar{S}-i}\right)
\partial_\m\bar{S}+\left(\frac{\bar{S}+i}{S+i}\right)\partial_\m S\right]
\, . \label{Q}
\ee

The field  equations  as follows from the action (\ref{d3}) are
\bea
R_{\m\n}&=&
\frac{\partial_\m S\partial_\n \bar{S}}{(S-\bar{S})^2}
+\frac{\partial_\n S\partial_\m \bar{S}}{(S-\bar{S})^2}\,
\\
0&=& \frac{1}{\sqrt{-g}}\partial_\m\left(g^{\m\n}\sqrt{-g}\frac{1}{(S-
\bar{S})^2}\partial_m S\right)+
\frac{\partial_\m S\partial^\m \bar{S}}{(S-\bar{S})^3}\, . \label{eqS}
\eea
We are looking for supersymmetric solutions of the form
\be
ds^2=-dt^2+e^{\rho(z,\bar{z})} dzd\bar{z} \, ,\label{metr}
\ee
and, as usual, the conditions for unbroken supersymmetry are
\bea
\delta\lambda=0&, &\delta\psi_\m=0 \, .
\eea
The field equations for the $S$-field in the background of eq.(\ref{metr}) is
\be
\partial\bar{\partial}S+ 2\frac{\partial S\bar{\partial}S}
{(S-\bar{S})}=0 \, ,
\ee
and it is obvious that  it is solved  by holomorphic
(or anti-holomorphic) functions $S=S(z)$ $\left(S=S(\bar{z})\right)$.
In this case, the $\delta\lambda=0$ condition is satisfied if
\be
\sigma^3\epsilon=\epsilon\, . \label{se}
\ee
Moreover, if $\epsilon$ satisfies eq.(\ref{se}), 
the integrability condition for holomorphic S-field 
of $D_\m\epsilon=0$ turns out to be
\be
\partial\bar{\partial}\rho=
\frac{\partial S\bar{\partial}\bar{S}}
{(S-\bar{S})^2}\, . \label{rho}
\ee
This equation coincides with the Einstein equations in
eq.(\ref{eqS}) and thus, it is the only one which remains to be
solved.  However, before doing that, let us consider the energy of
these configurations which is given by
\be
E=-\frac{i}{2}\int d^2z
\frac{\partial S\bar{\partial}\bar{S}}
{(S-\bar{S})^2}\, . \label{ene}
\ee
Since the $S$-field belongs to the upper half plane, the energy in
eq.(\ref{ene}) is infinite. In order to find finite energy solutions
one has to restrict $S$ to the fundamental domain ${\cal F}$
of PSL(2,Z)~\cite{g}. Thus, $S$ has discontinuous jumps
done by PSL(2,Z) transformations  $S\rightarrow S+1$
as we go around the source.  These jumps as well as the holomorphicity
require that near the sources
\be
S\simeq \frac{1}{2\pi i}\ln z \, .
\ee
The energy in this case is indeed finite and in particular
\be
E=\frac{\pi}{6}N \, , \label{ener}
\ee
where $N$ is the number of  times
the z-plane covers the fundamental domain ${\cal F}$.

Turning now to eq.(\ref{rho}), one may easily verify that it is solved by
\be
\rho=S_2 |h(S)|^2 \, ,
\ee
where $S_2=ImS$ and
$h(S)$ is an arbitrary integration function. It can be specified
by demanding modular invariance and nowhere vanishing metric. These
two conditions give
the supersymmetric solution~\cite{g}
\be
e^{\rho}=S_2\eta(S)^2\bar{\eta}(S)^2
|\prod_{i=1}^{N}(z-z_i)^{-1/12}|^2 \, , \label{Eins}
\ee
where $\eta(S)$ is the Dedekind's $\eta$-function.
The asymptotic form of the space-time is then
\be
ds^2\sim -dt^2+|z\bar{z}|^{-N/12}dzd\bar{z} \, ,
\ee
so that the space-time develops a deficit angle $\delta=\pi N/6$.
The constraint eq.(\ref{se}) breaks  half of the supersymmetries 
and although the   space-time is conical, we managed to find a Killing spinor
because the spin connection is  exactly canceled by the
U(1) gauge field of eq.(\ref{Q}). This possibility has to be anticipated to the
one in [3] where an Aharonov-Bohm phase due to the gauge field cancels the
phase due to the spin connection. 

There exist  now Golstone fermions for the
broken supersymmetry which are the zero modes
\bea
\delta\lambda&=&-\frac{1}{S_2}\left(\frac{\bar{S}-i}{S+i}\right)
\gamma^\m\partial_\m S
\epsilon^{\ast}\, , \nonumber \\
\delta \psi_z&=&D_z \epsilon\, , \nonumber \\
\delta \psi_{\bar{z}}&=&D_{\bar{z}} \epsilon\, , \label{supa}
\eea
where now the spinor $\epsilon$ satisfies
\be
\sigma^3\epsilon=-\epsilon\, . \label{see}
\ee
In order the supersymmetry parameter $\epsilon$ to generate the zero modes,
it must have the asymptotic behaviour 
\be
\epsilon\rightarrow e^{-\rho/2}\epsilon_0 \, , 
\ee
at spatial infinity. In this case, however, the norm of the zero mode gets a
divergent contribution
\be
\int d^2z\delta\psi_z^\ast\delta\psi_z e^{-\rho}
\sim\int d^2z\frac{1}{z\bar{z}} \, ,
\ee
and thus, it does not appear in the physical Hilbert space. 

\section{Conclusions}

We discussed above solitonic solutions describing
 zero-branes in 2+1 dimensions which preserve
half of the supersymmetries. Massive particles in
planar gravity produce a deficit angle which also breaks the other half of
the supersymmetries. Here 
we have seen that in the particular supersymmetric theory in 2+1 dimensions 
we have considered, it was possible to define Killing spinors and consequently,
the unbroken supersymmetry  survives despite the conical structure. 
The Golstone fermions of the broken supersymmetry however, are not normalizable
and thus do not fill any representation of the unbroken one. On the other hand,
in the vacuum all supersymmetries are unbroken and the cosmological constant
vanishes. We see here a concrete example of Witten's observation for the
connection between the vanishing of the cosmological constant and the
bose-fermi degeneracy of the excited states.



\begin{thebibliography}{99}
\bibitem{1} H.B. Nielsen and P. Olesen, Nucl. Phys. B 61 (1973)45.
\bibitem{2} A. Comtet and G.W. Gibbons, Nucl. Phys. B 299 (1988)719.
\bibitem{3} K. Becker, M. Becker and A. Strominger,
Phys. Rev. D 51 (1995)6603. 
\bibitem{4} A. Staruszkiewicz, Acta Physica Pol. 24 (1963)735;
G. Clement, Nucl. Phys. B 114 (1976)437; P. Collas, 
Amer. J. Phys. 45 (1977)833; S. Giddings, J. Abbott and K. Kuchar,
Gen. Rel. Grav 16 (1984)751.
\bibitem{5} S. Deser, R. Jackiw and G. 't Hooft, 
Ann. Phys. 152 (1984)220.
\bibitem{5'}M. Henneaux, Phys. Rev. D 29 (1984)2766.
\bibitem{6} P.S. Howe, J.M. Izquierdo, G. Papadopoulos and P.K. Townsend,
Nucl. Phys. B 467 (1996)183.
\bibitem{w}E. Witten, {\em ``Is Supersymmetry Really Broken?"},
preprint IASSNS-HEP-94-72, hep-th/9409111.
\bibitem{nishino} H. Nishino, Phys. Lett. B 370 (1996)65.
\bibitem{deserab}L.F. Abbott and S. Deser,
Nucl. Phys. B 195 (1982)76; Phys. Lett. B 116 (1982)259.
\bibitem{teit}C. Teitelboim,
Phys. Lett. B 69 (1977)240; Phys. Rev. D 29 (1984)2763.
\bibitem{color} A. Abouelsaood, Phys. Lett. B 125 (1983)467;
Nucl. Phys. B 226 (1984)309;A. Balachandran, G. Marmo, M. Mukunda,
J. Nilsson, E. Sudarshan and F. Zaccaria,
Phys. Rev. lett. 50 (1983)1553; S. Coleman and P. Nelson,
Nucl. Phys. B 237 (1984)1.
\bibitem{deser}S. Deser, Class. Quantum
Grav. 2 (1985)489.
\bibitem{a} M. Gell--Mann and B. Zwiebach, Phys. Lett. B 147 (1984)111.
\bibitem{b} G.W. Gibbons, M.B. Green and M.J. Perry,
Phys. Lett. B 370 (1996)37. 
\bibitem{g} B.R. Greene, A. Shapere, C. Vafa and S.T. Yau,
Nucl. Phys. B 337 (1990)1.
\bibitem{d} J.H. Schwarz, Nucl. Phys. B 226 (1983)269.
\end{thebibliography}
\end{document}